\documentclass[12pt]{iopart}

\usepackage{iopams}
\newtheorem{lemma}{Lemma}
\newtheorem{proposition}{Proposition}

\newcommand{\be}{\begin{equation}}
\newcommand{\ee}{\end{equation}}
\def\tr{\mathop{\rm tr}\nolimits}

\def\dif{{\rm d}}

\def\ci{{\rm i }}
\usepackage[latin1]{inputenc}
\usepackage[english]{babel}
\usepackage{graphics}
\usepackage{graphicx}

\begin{document}

\title[Obtaining the multiple Debever null directions]
{Obtaining the multiple Debever null directions}

\author{Juan Antonio S\'aez$^1$, Salvador Mengual$^2$\footnote{Author to whom any correspondence should be addressed.} and Joan Josep Ferrando$^{2,3}$}

\address{$^1$\ Departament de Matem\`atiques per a l'Economia i l'Empresa,
Universitat de Val\`encia, E-46022 Val\`encia, Spain}

\address{$^2$\ Departament d'Astronomia i Astrof\'{\i}sica, Universitat
de Val\`encia, E-46100 Burjassot, Val\`encia, Spain}

\address{$^3$\ Observatori Astron\`omic, Universitat
de Val\`encia, E-46980 Paterna, Val\`encia, Spain}

\ead{juan.a.saez@uv.es; salvador.mengual@uv.es; joan.ferrando@uv.es}

\begin{abstract}
The explicit expression of the multiple Debever null directions of an algebraically special spacetime are obtained in terms of the electric and magnetic parts of the Weyl tensor. An algorithm for the determination of the Petrov-Bel type and the algorithm to obtain the multiple Debever null directions are implemented as functions of a new package of \textit{xAct}, a \textit{Mathematica} suite of packages for tensor manipulations. These functions are applied to two examples. 
\end{abstract}
%
%

\vspace{1mm}

\noindent

\vspace{3mm}

\section{Introduction}

Here we follow the notation and terminology introduced and explained in \cite{FMS-Weyl}. Let $(V_4, \, g)$ be an oriented and time-oriented spacetime of signature $\left\lbrace -, +, +, + \right\rbrace$, ${\cal W} = \frac{1}{2} (W- \ci *W)$ the self-dual Weyl tensor and ${\cal G} = \frac{1}{2} (G -\ci \eta)$ the canonical metric on the space of self-dual 2-forms (bivectors), where $G = \frac{1}{2} \, g \wedge g$, and $\eta$ is the metric volume element. The algebraic classification of the Weyl tensor W can be obtained \cite{Petrov, bel-3} by studying the traceless linear map defined by the self-dual Weyl tensor on the space of bivectors. An alternative viewpoint consists of studying the relative positions between the `null cones' determined by the canonical metric and the Weyl tensor as a quadratic form on the bivectors \cite{Debever-56}. These `null cones' cut, generically, on four null bivectors ${\cal H}$ that are the solutions of the equations ${\cal G}({\cal H}, {\cal H})=0$, ${\cal W}({\cal H}, {\cal H})=0$. We call them {\em Debever null bivectors} \cite{FMS-Weyl}. The fundamental vector $\ell$ of each one defines a null direction on the spacetime, which is usually called {\em Debever null direction} \cite{Debever}.

Petrov-Bel types N, III and II admit a single multiple Debever null direction which is quadruple, triple and double, respectively. These three types correspond to the Bel radiative gravitational fields, and this multiple Debever direction is called the {\em fundamental direction} of the gravitational field \cite{bel-3}. Petrov-Bel type D admits two double Debever directions, and in type I there is no multiple Debever direction.  

The underlying geometry associated with each Petrov-Bel type was widely analyzed in \cite{FMS-Weyl}, a paper where we also offered the algorithm to determine all these geometric elements defined by the Weyl tensor, and in particular to obtain the Debever null directions. In Sec. 2 of the present paper, we offer an alternative approach based on the Bel-Robinson tensor \cite{bel-3} to determine the fundamental direction of a Bel radiative gravitational field. This method enables us to give them as explicit concomitants of the Weyl tensor. 

As usual in Relativity, an observer is defined by a unit time-like vector $u$. The electric and magnetic parts of the Weyl tensor relative to an observer, $E$ and $H$ respectively, play a crucial role in the analysis of gravitational radiation, since they represent the locally free curvature. In this context, Ricci and Bianchi identities, which are fundamental equations of this approach, are expressed in terms of such quantities \cite{Maartens_nou}. The use of the 3+1 tetrad formalism \cite{vanElst_Uggla} to analyze these equations has been useful in studying gravitational radiation and in analyzing some cosmological models (see \cite{Maartens, Ellis_Maartens} and references therein). Furthermore, the decomposition of the Weyl tensor into its electric and magnetic parts is also used in several works where the initial data of a given spacetime is characterized (see, for example \cite{Alfonso-ID-1,Alfonso-ID-2,Alfonso-ID-3}). Thus, it is important to obtain explicit expressions of geometric data such as the multiple Debever null directions in terms of $E$, $H$ and the observer $u$. In Sec. 2 we also get such expressions for both, the Bel radiative fields and the type D space-times.

Our results offer an IDEAL (Intrinsic, Deductive, Explicit and ALgorithmic) way to determine the multiple Debever directions. An approach of this kind is an essential tool in performing the IDEAL characterization of a specific solution of the Einstein field equations or other physical fields. The interest of these IDEAL studies has been widely analyzed elsewhere (see \cite{fs-SSST, fs-Szafron-Ideal} and references therein) and our results in this paper could be useful in further work. For example, if we know the explicit expression of the multiple Debever direction in terms of the Weyl tensor (or its electric and magnetic parts), $\ell = \ell(W)$, then the spacetimes admitting an non-twisting geodesic multiple Debever direction will be characterized by the equation $\dif \ell(W) \wedge \ell(W) = 0$, a covariant condition that can be tested if we know the metric tensor in an arbitrary coordinate system. Of course, the determination of the multiple Debever directions for an algebraically special gravitational field can be useful for studying gravitational radiation as well as for constructing invariant frames which enable us to analyze the intrinsic properties of the metric tensor.

Another compelling aspect of IDEAL determinations is that their algorithmic structure enables straightforward implementation into computer algebra programs.  Computer algebra is a term used to refer to the theory and implementation of computer programs to perform the symbolic manipulations and calculations usual in mathematics. It has become an essential tool in gravity research \cite{MacCallum}. \textit{xAct} is an example of such computer algebra programs. It is actually a collection of packages for \textit{Mathematica} and it has attracted a fair amount of users, as can be seen from the extensive bibliography of papers and theses that have used it (see the articles section in \cite{xAct}). In Sec. 3, we show how the implementation of the algorithms presented in this paper on \textit{xAct} enable us to obtain the Petrov-Bel type and multiple Debever null directions of some paradigmatic metrics as examples. In further works, we plan to implement other algorithms we have developed to characterize some families of solutions (see \cite{fs-Szafron-Ideal} and references therein).

Of course, there exist other algorithms that give the geometric data considered in this paper (such as the Cartan-Karlhede algorithm \cite{Cartan-Karlhede}) and some of them are also implemented in other computer algebra programs (see \cite{PollneyI, PollneyII, PollneyIII, MacCallum} and references therein). However, their approach is quite different from the one presented in this paper. In Sec. 4, we discuss how the different methods compare and explain why it is convenient to have all of them available.

\section{Algorithmic determination of the multiple Debever null directions}

The interest of the Bel-Robinson tensor in analyzing radiative gravitational states has been widely reported (see \cite{fsRS, seno, garcia-parrado, fsRS-b, wylleman, fran} and references therein). Here, we use it as a mathematical tool to obtain the fundamental direction of the Bel radiative fields. In terms of the self-dual Weyl tensor ${\cal W}$, the Bel-Robinson tensor takes the expression \cite{fsBR-1, fsBR-2} (symbol $\, \bar{ } \,$ denotes the complex conjugate):
\begin{equation} \label{BR}
T_{\alpha \beta \mu \nu} \equiv {{{ {{\cal W}_{\alpha}}^{\lambda}}_\beta}^{\rho}} \ \bar{{\cal W}}_{ \lambda \mu  \rho \nu} = \bar{{\cal W}}_{\alpha \lambda  \beta  \rho} \ {{{ {{\cal W}^{\lambda}}_{\mu}}^\rho}_{\nu}} \, .
\end{equation}

	\subsection{Fundamental direction of a Bel radiative gravitational field}

A null bivector ${\cal H}$ can be written as ${\cal H} = \ell \wedge m$, where $\ell$ is a real null vector (which defines the fundamental direction of ${\cal H}$) and $m$ is a complex null vector such that $\ell^2 = m^2 = \ell \cdot m = 0$, $m \cdot \bar{m} = 1$. Then, we obtain that ${\cal H} \cdot \bar{\cal H} = -\ell \otimes \ell$, where $\, \cdot \,$ denotes the contraction of adjacent indices. There is a one-to-one correspondence between a null bivector ${\cal H}$ and the self-dual {\em Weylian tensor} (double two-form with the same symmetries as the Weyl tensor) given by:
\begin{equation} \label{Omega}
\Omega \equiv  {\cal H} \otimes {\cal H}   \, ,
\end{equation}
which is characterized by condition $\Omega^2 = 0$, where $(\Omega^2)_{\alpha \beta}^{\ \ \lambda \mu} = \frac12 \Omega_{\alpha \beta}^{\ \ \sigma \rho} \Omega_{\sigma \rho}^{\ \ \lambda \mu}$. If $P$ is the Bel-Robinson-like tensor associated with the Weylian tensor $\Omega$, $P_{\alpha \beta \mu \nu} \equiv {{{{\Omega_{\alpha}}^{\lambda}}_\beta}^{\rho}} \ \bar{\Omega}_{ \lambda \mu  \rho \nu}$, we obtain:
\begin{equation} \label{P}
P =( {\cal H} \cdot \bar{\cal H}) \otimes ({\cal H} \cdot \bar{\cal H}) = \ell \otimes \ell \otimes \ell \otimes \ell \, .
\end{equation}

On the other hand, the algebraic study of the self-dual Weyl tensor for Petrov-Bel types N, III, II (radiative gravitational fields) presented in [1], and taking into account that the multiple Debever bivector ${\cal H}$ is determined up to a factor, we obtain the following result:

\begin{lemma} \label{lemma-1}
Let ${\cal W}$ and ${\cal H}$ be the self-dual Weyl tensor and the multiple Debever bivector of a Bel radiative gravitational field, and $\Omega \equiv  {\cal H} \otimes {\cal H}$. Then: (i) $\Omega \equiv {\cal W}$ if the Weyl tensor is of type N, (ii) $\Omega \equiv -{\cal W}^2$ if the Weyl tensor is of type III, and (iii) $\Omega \equiv {\cal W}^2 + \rho {\cal W} - 2 \rho^2
{\cal G} , \ \rho = -\frac{\Tr {\cal W}^3}{\Tr {\cal W}^2}$, if the Weyl tensor is of type II.
 \end{lemma}
 In the previous lemma, and hereinafter, for any double two-form $\cal F$ we write $\textrm{Tr} \, {\cal F} = \frac12 {\cal F} ^{\, \alpha \beta} _{\ \ \alpha \beta}$. From this lemma and expressions (\ref{Omega}) and (\ref{P}), we obtain:
\begin{proposition} \label{proposition-1}
The fundamental direction $\ell$ of a Bel radiative gravitational field can be obtained as
\begin{equation} \label{ell}
\ell_{\nu} \propto w^{\alpha} w^{\beta} w^{\mu} P_{\alpha \beta \mu \nu} \, ,
\end{equation}
where $w$ is an arbitrary time-like vector, and where $P$ is the Bel-Robinson-like tensor associated with the Weylian tensor $\Omega$,
\begin{equation} \label{P-b}
P_{\alpha \beta \mu \nu} \equiv {{{ {\Omega_{\alpha}}^{\lambda}}_\beta}^{\rho}} \ \bar{\Omega}_{ \lambda \mu  \rho \nu}  \, ,
\end{equation}
where
\begin{itemize}
\item[(i)]
$\Omega \equiv {\cal W}$, if the Weyl tensor is of type N.
\item[(ii)]
$\Omega \equiv -{\cal W}^2$, if the Weyl tensor is of type III.
\item[(iii)]
$\Omega \equiv {\cal W}^2 + \rho {\cal W} - 2 \rho^2
{\cal G} ,  \  \rho = -\frac{\Tr {\cal W}^3}{\Tr {\cal W}^2}$, if the Weyl tensor is of type II.
 \end{itemize}
 \end{proposition}

If $E$ and $H$ are the electric and magnetic parts of the Weyl tensor with respect to an observer $u$, the {\em Petrov matrix} relative to $u$ is ${\cal Q}= E- \ci H$, and it can be obtained as:
\begin{equation}\label{Q-W}
{\cal Q}_{\beta \nu} = 2 u^{\alpha} u^{\mu} \, {\cal W}_{\alpha \beta \mu \nu} \, .
\end{equation}
Conversely, in terms of $u$, and ${\cal Q}$, we have:
\begin{equation} \label{W-Q}
\hspace{-5mm}2 {\cal W} =  - u \wedge {\cal Q} \wedge u + *(u \wedge {\cal Q} \wedge
u) * + \ci  *(u \wedge {\cal Q} \wedge u) + \ci \, (u \wedge {\cal Q} \wedge u)*  ,
\end{equation}
where, for a two-tensor $B$ and a vector $x$, $(x \wedge B)_{\alpha \beta \gamma} = x_\alpha  B_{\beta \gamma} - x_\beta B_{\alpha \gamma}$ and $(B \wedge x)_{\alpha \beta \gamma} = B_{\alpha \beta} x_{ \gamma} - B_{\alpha \gamma}x_\beta$, 
and for  a double two-form ${\cal A}$, $*{\cal A}_{\alpha \beta \gamma \delta} = \frac12 \eta_{\alpha \beta \lambda \mu} {\cal A}^{\lambda \mu}_{\ \ \gamma \delta}$ and ${\cal A}*_{\alpha \beta \gamma \delta} = \frac12 \eta_{\gamma \delta \lambda \mu} {\cal A}_{\alpha \beta}^{\ \ \lambda \mu}$. Expression (\ref{W-Q}) is the self-dual version of a formula that can be easily shown by applying twice the known expression that gives a 2-form in terms of an observer and its relative electric and magnetic fields \cite{Hawking, Coll-Tesi}.

Similarly, we can associate a Petrov matrix ${\cal  P}$ with the Weylian tensor $\Omega$, ${\cal P}_{\beta \nu} = 2 u^{\alpha} u^{\mu} \, \Omega_{\alpha \beta \mu \nu}$. Lemma \ref{lemma-1} shows that, in the Bel radiative types, a dependence between the self-dual Weyl tensor ${\cal W}$ and the Weylian tensor $\Omega$ exists. Then, a straightforward calculation shows that definition (\ref{Q-W}) and expression (\ref{W-Q}) induce a dependence between their associated Petrov matrices: ${\cal P} = {\cal Q}$ in type N, ${\cal P} = -{\cal Q}^2$ in type III, and ${\cal P} = \rho {\cal Q} + 2 \rho^2 \gamma - {\cal Q}^2 ,  \ \gamma = g + u \otimes u,  \ \rho=\frac{\tr {\cal Q}^3}{\tr {\cal Q}^2}$, in type II.

On the other hand, if we use (\ref{P-b}) to calculate $P$, and we substitute this $P$ in (\ref{ell}) and consider the arbitrary time-like vector $w$ equal to the observer $u$, we arrive to the following.
\begin{proposition} \label{proposition-2}
In terms of the observer $u$ and the Petrov matrix ${\cal Q}= E- \ci H$, the fundamental direction $\ell$ of a Bel radiative gravitational field can be obtained as
\begin{equation} \label{ell-b}
\hspace{-23mm}
 \ell_\nu ={\bar{\cal P}}^{\alpha \beta} [ {\cal P}_{\beta \alpha} u_{\nu} + \ci \,
 {\cal P_\beta}^\mu \, {\eta}_{\alpha \mu \nu \lambda} u^\lambda]  , \ \    \quad   \left[\ell  = \tr({\bar{\cal P}} \cdot {\cal P}) u  + 2 \, \ci \, *\!({\bar{\cal P}} \cdot {\cal P})(u)\right] ,
 \end{equation}
where
\begin{itemize}
\item[(i)]
${\cal P} \equiv {\cal Q}$ if the Weyl tensor is of type N.
\item[(ii)]
${\cal P} \equiv - {\cal Q}^2$ if the Weyl tensor is of type III.
\item[(iii)]
${\cal P} \equiv \rho {\cal Q} + 2 \rho^2
\gamma - {\cal Q}^2 ,  \ \gamma = g + u \otimes u,  \ \rho=\frac{\tr
{\cal Q}^3}{\tr {\cal Q}^2}$, if the Weyl tensor is of type II.
 \end{itemize}
\end{proposition}
In the above proposition and in what follows, for a 2-tensor $B$ and a vector $x$, $B(x)$ denotes the one-form $B(x)_\alpha = B_{\alpha \beta} x^\beta$.

	\subsection{Null principal directions of a type D Weyl tensor}

A type D Weyl tensor takes the canonical expression \cite{FMS-Weyl}
\begin{equation} \label{D-can}
{\cal W}  = 3 \rho \, {\cal U} \otimes {\cal U} + \rho \, {\cal G} \, ,  \qquad  {\cal U} = \frac{1}{\sqrt{2}}(U - \textrm{i} *U) \, , \qquad \rho = -\frac{\Tr {\cal W}^3}{\Tr {\cal W}^2} ,
\end{equation}
where ${\cal U}$ is the simple (non null) eigen-bivector, and $\rho$ the double eigenvalue. The simple and unitary real two-form $U$ is the volume element of the time-like {\em principal plane} of a type D Weyl tensor. The two double null Debever directions $\ell_{\pm}$ are the {\em null principal directions} of $U$. Then, on the one hand, in \cite{FMS-Weyl} it is shown that:

\begin{proposition} \label{proposition-3}
Let ${\cal W}$ be a type D Weyl tensor. Then, the double Debever null directions $\ell_\pm$ can be obtained as
\begin{equation}
\ell_\pm \propto \left[ U^2 \pm U \right] (x) \, ,
\end{equation}
where x is an arbitrary timelike future-pointing vector and $U$ can be obtained as
\begin{equation} \label{TypeD-directions}
U = \frac{1}{\sqrt{2}} ({\cal U} + \bar{\cal U}) \, , \qquad {\cal U} \equiv \frac{{\cal A} ({\cal X})}{\sqrt{- {\cal A} ^2 ({\cal X} , {\cal X})}} \, , \qquad {\cal A} \equiv {\cal W} - \rho \, {\cal G} \, ,
\end{equation}
with ${\cal X}$ an arbitrary bivector such that ${\cal A} ({\cal X}) \neq 0$.
\end{proposition}
In the above proposition, for a double two-form ${\cal A}$ and a bivector ${\cal X}$, ${\cal A}({\cal X})$ denotes the bivector ${\cal A}({\cal X})_{\alpha \beta} =\frac12 {\cal A}_{\alpha \beta \lambda \mu} {\cal X}^{\lambda \mu}$. The bivector ${\cal A}$ given in (\ref{TypeD-directions}) is the projector on the eigen-bivector ${\cal U}$. Therefore, ${\cal A}({\cal X})$ determines this eigen-direction as long as ${\cal X}$ is not orthogonal to it.

On the other hand, the orthogonal projection of each observer $u$ defines a time-like direction $e_0$ on the plane $U$ and, generically, a direction $e_2$ on the plane $*U$. If we complete  an orthonormal  frame $\{e_\alpha \}$ taking $e_1$ in $U$ and $e_3$ in $*U$, we have
\begin{equation}\label{observer}
u= \cosh \phi \, e_0  + \sinh \phi \, e_2 \,, \qquad U=e_0 \wedge e_1 \,
, \quad *U= e_3 \wedge e_2 \, .
\end{equation}
%
Then, the two Debever directions are $\ell_{\pm} \propto e_0 \pm e_1$. From (\ref{D-can}),
(\ref{observer}), and the expression of the Petrov matrix (\ref {Q-W}), we obtain:
\be \label{P-c}
\hspace{-10mm} {\cal R} \equiv \frac{1}{3 \rho}{\cal Q}=   \cosh^2 \! \phi \, e_1 \otimes e_1 -  \sinh^2 \! \phi \, e_3 \otimes e_3
- \frac{\ci}{2}  \sinh  2 \phi \, e_1 \stackrel{\sim}{\otimes} e_3 - \frac{1}{3} \gamma \, ,
\ee
where $\gamma=g + u \otimes u$, and for two vectors $x$, $y$, $x \stackrel{\sim}{\otimes} y = x \otimes y + y \otimes x$. From expression (\ref{P-c}), we can compute the vector $v_0 \equiv \cosh \phi \, e_0$ and the projector on the direction $e_1$, $S  \equiv  \cosh^2 \phi \, e_1 \otimes e_1$, in terms of the tensor ${\cal R}$. Note that if $S(w)$ does not vanish, it determines the direction of $e_1$. Then, we obtain:
\begin{proposition} \label{proposition-4}
In terms of the observer $u$ and the Petrov matrix ${\cal Q}$, the null principal directions $\ell_{\pm}$ of a type D Weyl tensor can be obtained as
\begin{equation} \label{ell+-}
\ell_{\pm} = v_0 \pm  v_1 \,  ,
\end{equation}
where 
\begin{equation} \label{e0e1}
\hspace{-17mm} v_0 \equiv \cosh \phi \, e_0 = \cosh^2  \phi\, u- \frac{\ci}{\sqrt{\zeta}} \,
*({\bar{\cal R}} \cdot {\cal R})(u) \, , \quad v_1 \equiv \cosh \phi \, e_1 = \frac{S(w)}{\sqrt{S(w,w)}} \, ,
\end{equation}
$w$ being an arbitrary vector such that $S(w) \not=0$, and with
 \begin{eqnarray} \label{P-S}
\hspace{-22mm} {\cal R} \equiv \frac{1}{3 \rho}{\cal Q}  \, ,  \quad S \equiv \frac{1}{4}(1 + \frac{2}{3\sqrt{\zeta}})({\cal R}+ \bar{\cal R}) +\frac{1}{4 \sqrt{\zeta}}   ({\cal R} \cdot \bar{\cal R} + \bar{\cal R} \cdot {\cal R} )   + \frac16 (1 + \frac{1}{3\sqrt{\zeta}}) \gamma \, ,  \\
\hspace{-22mm} \rho \equiv \frac{\tr {\cal Q}^3}{\tr {\cal Q}^2} \, , \quad \zeta \equiv \tr[{\cal R} \cdot \bar{\cal R}] + \frac13 \, , \quad \cosh^2 \phi \equiv \frac12 (1+ \sqrt{\zeta}) \, , \quad   \gamma=g + u \otimes u    \, . \label{alpha-z-a}
\end{eqnarray}
 \end{proposition}
Most of the expressions that we use for getting the above proposition have also been recently obtained in \cite{wylleman}.

	\subsection{An algorithm to determine the Petrov-Bel type}

The above results show that the explicit expression of the multiple Debever null directions depends on the Petrov-Bel type.
In order to obtain this type, we can make use of any of the algorithms that exist in the literature. Below, we provide a flow chart that shows an algorithm with conditions only involving the Petrov matrix ${\cal Q}$. It is an adapted version of that presented in \cite{FMS-Weyl}, and it can be deduced considering expression (\ref{W-Q}). The input data are the spatial metric $\gamma=g + u \otimes u$, the Petrov matrix ${\cal Q} \not=0$, and the scalars $a \equiv \tr {\cal Q}^2$ and $b \equiv -\tr {\cal Q}^3$. The case ${\cal Q}=0$ leads to a vanishing Weyl tensor (type O).

\vspace{-0.5cm}

 \hspace*{-2mm} \setlength{\unitlength}{0.9cm} {\small
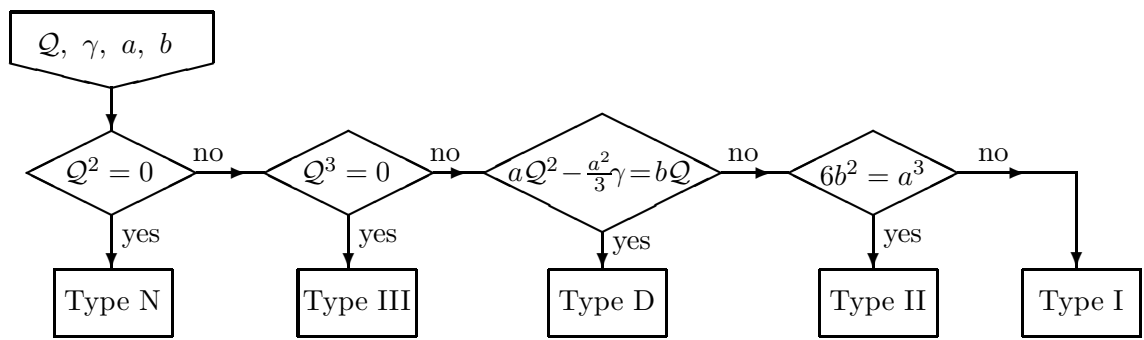
\noindent
\begin{picture}(0,13)


\thicklines \put(2,11){\line(-4,-1){1.5}}
 \put(-1,11){\line(4,-1){1.5}}
\put(-1,11.75){\line(0,-1){0.75}} \put(2,11.75){\line(-1,0){3}}
\put(2,11.75){\line(0,-1){0.75}} \put(-0.9,11.15){$ \ \ {\cal Q} , \
\gamma, \ a , \ b  $}

\put(0.5,10.65){\vector(0,-1){0.65}}

\put(0.5,10){\line(-2,-1){1.25}} \put(0.5,10){\line(2,-1){1.25}}
\put(0.5,8.75){\line(2,1){1.25}} \put(0.5 ,8.75){\line(-2,1){1.25}}
\put(-0.20,9.25){${\cal Q}^2 =0 $}

\put(4,10){\line(-2,-1){1.25}} \put(4,10){\line(2,-1){1.25}}
\put(4,8.75){\line(2,1){1.25}} \put(4,8.75){\line(-2,1){1.25}}
\put(3.3,9.25){${\cal Q}^3 =0 $}

\put(7.75,10.25){\line(-2,-1){1.75}}
\put(7.75,10.25){\line(2,-1){1.75}} \put(7.75,8.5){\line(2,1){1.75}}
\put(7.75,8.5){\line(-2,1){1.75}} \put(6.35,9.25){$b {\cal Q}\!= \!\frac{a^2}{3} \!\gamma\!-\!a {\cal Q}^2$}


\put(11.75,10){\line(-2,-1){1.25}} \put(11.75,10){\line(2,-1){1.25}}
\put(11.75,8.75){\line(2,1){1.25}}
\put(11.75,8.75){\line(-2,1){1.25}} \put(10.96,9.2){$6 b^2 = a^3 $}


\put(4,8.75){\vector(0,-1){0.8}} \put(0.5,8.75){\vector(0,-1){0.8}}

\put(11.75,8.75){\vector(0,-1){0.8}}

 \put(7.75,8.5){\vector(0,-1){0.55}}

 \put(1.75,9.38){\vector(1,0){0.75}}\put(2.5,9.38){\line(1,0){0.25}}

\put(5.25,9.38){\vector(1,0){0.5}}\put(5.75,9.38){\line(1,0){0.25}}

\put(9.5,9.38){\vector(1,0){0.75}}
\put(10.25,9.38){\line(1,0){0.25}}

\put(13 ,9.38){\vector(1,0){1}}\put(14 ,9.38){\line(1,0){0.76}}
 \put(14.75,9.37){\vector(0,-1){1.4}}

\put(3.25,6.95){\line(1,0){1.75}} \put(3.25,6.95 ){\line(0,1){1}}
\put(5,7.95 ){\line(-1,0){1.75}} \put(5,7.95 ){\line(0,-1){1}}
\put(3.32,7.35){Type III}

\put(-0.35,6.95){\line(1,0){1.75}} \put(-0.35,6.95){\line(0,1){1}}
\put(1.4,7.95 ){\line(-1,0){1.75}} \put(1.4,7.95 ){\line(0,-1){1}}
\put(-0.2,7.35){Type N}


\put(6.95,6.95){\line(1,0){1.75}} \put(6.95,6.95 ){\line(0,1){1}}
\put(8.7,7.95 ){\line(-1,0){1.75}} \put(8.7,7.95 ){\line(0,-1){1}}
\put(7.1,7.35){Type D}

\put(10.95,6.95){\line(1,0){1.75}} \put(10.95,6.95){\line(0,1){1}}
\put(12.7,7.95 ){\line(-1,0){1.75}} \put(12.7,7.95){\line(0,-1){1}}
\put(11.1,7.35){Type II}

\put(13.95,6.95){\line(1,0){1.75}} \put(13.95,6.95){\line(0,1){1}}
\put(15.7,7.95 ){\line(-1,0){1.75}} \put(15.7,7.95){\line(0,-1){1}}
\put(14.19,7.35){Type I}

\put(11.9,8.35){yes} \put(7.9,8.25){yes}

\put(4.15,8.35){yes} \put(0.65,8.35){yes}

\put(13.3,9.55){no} \put(9.6,9.55){no} \put(5.2,9.55){no}
 \put(1.7,9.55){no}
\end{picture} }

\vspace{-5.5cm}

\section{Implementation of the algorithms on \textit{xAct}}

As already mentioned before, IDEAL determinations are particularly convenient for incorporation into a formal calculation computer program due to their algorithmic characteristics. An example of such a formal computational program is \textit{xAct}, a collection of \textit{Mathematica} packages for tensorial computer algebra \cite{xAct}. One of the authors of this paper (SM) is collaborating with A. Garc\'ia-Parrado on the development of an \textit{xAct} package named \textit{xIdeal}. This package incorporates IDEAL characterizations and other algorithms devised by our group. It will be published elsewhere upon completion, but its current unfinished version is available at \cite{xIdeal}.

In this section, we will use \textit{xIdeal} to determine the Petrov-Bel type and the multiple Debever null directions of two different metrics as examples.


	\subsection{An example of a radiative gravitational field}

The following metric is a particular case of a type II perfect fluid solution with a geodesic, shearfree, non-expanding multiple Debever null direction \cite{Kramer}:
\begin{equation}
	\hspace{-1.5cm} ds^2 = 2 F^{-2} (dx^2 + dy^2) - 2(dv + 2L \, dy)[dr + 2B \, dy + A(dv + 2L \, dy)] \, ,
\end{equation}
with $F^2 = \frac43 \kappa_0 x^3$, $A = -\kappa_0 x/2$, $B = \frac34 \sqrt{2}/x$  and  $L = \frac38 \sqrt{2}/(\kappa_0 x^2)$. It is a good example to check our functions since its Petrov-Bel type and null direction are already known. 

First, we check that, indeed, it is of Petrov-Bel type II with the $\mathtt{PetrovType}$ function. This function can determine the Petrov-Bel type of a given metric following two different algorithms: the one based on the self-dual Weyl tensor explained in \cite{FMS-Weyl} (by default) or the one shown at the end of the previous section. With both, we get that the considered metric is of type II. Then, we use the $\mathtt{DebeverNullDirections}$ function to obtain its multiple Debever null directions. Again, this function can follow the algorithm based on the self-dual Weyl tensor (by default) or the one based on the Petrov matrix obtained in the present paper. Regardless of the method, we get that the multiple Debever null vector has the direction of $\partial_r$ as reported in the literature.

\subsection{An example of type D}

A similar approach can be followed with Kerr-NUT metric. Kerr-NUT metric generalizes both Kerr and NUT vacuum solutions, and can be written as \cite{Kerr-Weir}
\begin{equation}
	\hspace{-2.4cm} ds^2 = - \frac{\alpha^2}{x^2 + y^2}(y^2 dt + dz)^2 + \frac{x^2 + y^2}{\alpha^2} dx^2 + \frac{\beta^2}{x^2 + y^2}(-x^2 dt + dz)^2 + \frac{x^2 + y^2}{\beta^2}dy^2 \, ,
\end{equation}
where
\begin{equation} \label{alpha i beta}
	\hspace{-1cm} \alpha^2 = p x^2 + \frac{k(3 - k^2)}{(1 + k^2)^3} x + s \, , \qquad \qquad \beta^2 = -p y^2 + \frac{3k^2 - 1}{(1 + k^2)^3}y + s \, .
\end{equation}
By setting $3k^2 = 1$ we recover the Kerr metric.

By using the $\mathtt{PetrovType}$ function, we get that the Kerr-NUT metric is of type D. It is worth remarking that, in our computations, we added the assumption that the coordinates satisfy $x^2 + y^2 > 0$. This helps the program to simplify some expressions and this, in turn, improves the computation speed. We explain this in more detail in the next section. For the same reason, we did not specify the expressions of the scalar functions $\alpha$ and $\beta$. Thus, the results we get here are valid for any $\alpha(x)$ and $\beta(y)$. However, only expressions (\ref{alpha i beta}) are solution of the vacuum Einstein equations.

In order to get the multiple Debever null directions of Kerr-NUT metric, we use the $\mathtt{DebeverNullDirections}$ function with the following extra assumptions: $y^4 \alpha^2 + x^4 \beta^2 > 0$ and $y^4 \alpha^2 - x^4 \beta^2 > 0$. These make sure that the function computes the quantity $\zeta$ defined in (\ref{alpha-z-a}) with the proper sign taking into account that we are considering the region outside the horizon. With that, we get that the two multiple Debever null directions of a Kerr-NUT metric are
\begin{equation}
 \ell_\pm^\mu = ( y^2 , \, \pm \, y^2 \, \alpha^2 , \, 0 \, , \, x^2 \, y^2 ) \, ,
\end{equation}
with $\alpha$ defined in (\ref{alpha i beta}). 

	\section{Other algorithms and discussion}
	
	As already mentioned at the beginning, there are other methods to obtain the geometric spacetime quantities discussed in this paper (see \cite{MacCallum} for an extensive review). One of the first such algorithms was the Cartan-Karlhede algorithm \cite{Cartan-Karlhede}, which was implemented by $\mathring{\rm{A}}$man in the CLASSI package \cite{Aman-Karlhede}, built on SHEEP \cite{SHEEP}. However, due to its lack of simplification methods and other limitations, others incorporated the same algorithm in more modern computer algebra programs. For instance, in \cite{PollneyI, PollneyII, PollneyIII}, the authors implemented it in GRTensor for \textit{Maple} using the spinor formalism. These algorithms need to obtain a canonical frame for the Weyl spinor as a first step, which involves solving a quartic equation. In some cases, the expression of this standard frame is too complicated or cannot even be found \cite{PollneyIII}. This is one of the reasons why alternative methods in which the frame is left completely arbitrary are necessary. This does not mean that free-frame methods will always provide a simpler solution faster than fixed-frame methods. Both approaches should be considered as complementary; depending on the spacetime considered and the frame in which it is expressed one approach (or none) may be preferable over the other.
	
Traditionally, tetrad based algorithms were thought to be more efficient than classical coordinate methods. However, as shown in \cite{Pollney}, a good simplification strategy is crucial to minimize the computing time. In the previous section, for example, we did not specify the expressions of $\alpha (x)$ and $\beta (y)$ for the Kerr-NUT metric, leaving them as arbitrary functions of the coordinates. We also added the assumption that the coordinates satisfy $x^2 + y^2 > 0$. This significantly improved the time spent simplifying some expressions. There is no universal rule to know which or how many assumptions are needed, but in some cases they are necessary to even get a result. In the previous section again, if we do not add the assumptions that $y^4 \alpha^2 + x^4 \beta^2 > 0$ and $y^4 \alpha^2 - x^4 \beta^2 > 0$, the multiple Debever null directions are not properly obtained. It is also worth mentioning that our functions require arbitrary vectors to compute some projections. The only theoretical restrictions on these arbitrary vectors is the non-nullity of some contractions as explained above. However, vectors that take simple expressions in the frame considered are obviously preferable.

As mentioned in the introduction, the algorithm presented in this paper, based on the Petrov matrix, may be interesting because it makes use of the electric and magnetic parts of the Weyl tensor relative to an observer, which in some contexts are the natural working variables. Regardless, this new algorithm can also be advantageous if it improves the computing time. Indeed, in the first example of the previous section, the Petrov matrix method requires less computing time than the algorithm presented in \cite{FMS-Weyl}, which is based on the self-dual Weyl tensor, whereas in the second example, only the method presented here provides a result in a reasonable time.


\ack We thank J M M Senovilla for his comments and A Garc\'ia-Parrado for sharing his knowledge about \textit{xAct}. This work has been supported by the Generalitat Valenciana Project CIAICO/2022/252 and the Plan Recuperaci\'on, Transformaci\'on y Resiliencia, project ASFAE/2022/001, with funding from European Union NextGenerationEU (PRTR-C17.I1). S.M. acknowledges financial support from the Generalitat Valenciana (grant CIACIF/2021/028).


\nopagebreak

\end{document}